\documentstyle[aps,pra]{revtex} 

\def\be{\begin{equation}}
\def\ee{\end{equation}}
\def\bea{\begin{eqnarray}}
\def\eea{\end{eqnarray}}
\def\L{{\cal L}_G}
\newcommand{\bra}[1]{\mbox{$\langle #1 |$}}
\newcommand{\ket}[1]{\mbox{$| #1 \rangle$}}
\newcommand{\proj}[1]{\mbox{$| #1 \rangle \! \langle #1 |$}}

\begin{document}
\draft

\title{Tight Bell inequality for $d$-outcome measurements correlations}

\author{Llu\'{\i}s Masanes }

\address{Dept. d'Estructura i Constituents de la Mat\`eria, Univ. Barcelona, 08028. Barcelona, Spain.}

\date{\today} 
\maketitle

\begin{abstract}
In this paper we prove that the inequality introduced by Collins, Gisin, Linden, Massar and Popescu \cite{Gisin} is tight, or in other words, it is a facet of the convex polytope generated by all local-realistic joint probabilities of $d$ outcomes. This means that this inequality is optimal. We also show that, for correlation functions generalized to deal with three-outcome measurements, the satisfyability of this inequality is a necessary and sufficient condition for the existence of a local-realistic model accounting for them. 
\end{abstract}


\section{Introduction}
Quantum Mechanics predicts that there are events in Nature exhibiting correlations which are not explainable in terms of local-realistic theories \cite{Bell,CHSH}. Those correlations can arise when measuring two or more separated systems which are in an entangled state. This fact is known as \emph{quantum nonlocality} and has been experimentally proven up to some loopholes \cite{Aspect}. Knowing which multipartite quantum states violate local-realism is a very important open problem. On the other hand, the sharing of non-local-realistic correlations between some parties is a useful resource for tasks like distributed computation \cite{cc} and secret communication \cite{cryptography}. Then, studying whether a given set of correlations are achievable with classical randomness or, contrary, they need of quantum entanglement, is an important issue.

It was shown in \cite{Peres} that the sets of local-realistic joint probability distributions are convex polytopes. There is one of these polytopes for each setting. By a setting we mean a fixed number of: parties, observables per party and number of outcomes per observable. These polytopes can be characterized by a finite number of linear inequalities that the joint probabilities of the correlated events must satisfy. These inequalities correspond to the \emph{facets} of the polytopes, here we call them \emph{tight Bell inequalities}. There are algorithms that find all the facets of a polytope, but the time they need for doing so grows very fast as the setting becomes less simple. Thus, obtaining these inequalities is in general a very hard problem, and it only has been completely solved in simple two-outcome settings \cite{Fine,Pitowsky}. In the case where correlation functions instead of joint probabilities are used, the problem is also completely solved for the setting consisting of an arbitrary number of parties each possessing two dichotomic observables, in a very mathematically-elegant way \cite{Werner}. In this paper we give a way for generalizing correlation functions to deal with $d$-outcome measurements. Using this instead of joint probabilities makes the problem of finding all the facets of the corresponding polytope numerically more feasible. We have done an algorithm that finds all the inequalities that characterize the set of local-realistic generalized correlation functions for any $d$. It turns out that for $d=2,3$, all the inequalities are equivalent to the CGLMP-inequality.

In the literature Bell inequalities exist which are not facets of correlation polytopes (\emph{non-tight Bell inequalities}), they are half-spaces that contain the polytope but do not lie in its frontier. In this sense we say that, {\em tight Bell inequalities} are the optimal detectors of non-local-realistic correlations. For most of the Bell inequalities there is no proof of tightness or non-tightness. In this paper it is shown that the CGLMP-inequality \cite{Gisin} is a \emph{tight Bell inequality}. In other words, the CGLMP-inequality is a facet of the polytope corresponding to the setting of two parties with two $d$-outcome observables per party. In fact, it is a family of equivalent facets of this polytope, but not all; thus, we do not completely solve the problem of characterizing all local-realistic joint probabilities for our setting, we just give a necessary condition for them.

This result is also important because often, the degree of violation of an inequality is used as a measure of how non-local a quantum state is. But this does not make sense unless a \emph{tight Bell inequality} is used, since otherwise, the measure has some inherent bias. 

Moreover, as it was shown in \cite{Acin}, the CGLMP-inequality (for $d=3$) is not maximally violated by the maximally entangled state when orthogonal measurements are done. This result becomes very surprising after knowing that this inequality is tight.
 
This paper is organized as follows. In section II we show that the set of all local-realistic correlations is a convex polytope, and we characterize it in terms of its generators. Section III contains the proof of tightness for the CGLMP-inequality. In section IV we simplify the problem in a way that makes possible to find numerically all the inequalities for $d=3$. Some conclusions are commented in section V. In the appendix, the two lemmas used in section II are proven.

\section{Characterization of the polytope}

In this section we justify the use of convex polytopes when studying local-realistic correlations. We also explain the dual description of a convex polytope: in terms of its generators on one side, and in terms of its facets on the other side. The setting that concerns us involves two parties: Alice and Bob. Alice can carry out two possible measurements, $A_1$ or $A_2$, and Bob can carry out $B_1$ or $B_2$. Each measurement has $d$ possible outcomes: $A_1, A_2, B_1, B_2 = 0, \ldots, d-1$. In the remaining of the paper the letters characterizing the measurements are also going to be used as the variables denoting their results. For each of the four experimental settings ($A_1 B_1, A_1 B_2, A_2 B_1, A_2 B_2$) there is a joint probability distribution of the outcomes:
\be
  P(A_a=k,B_b=s) \hspace{30pt} k,s=0,1,\ldots d-1  \hspace{30pt} a,b=1,2.
\label{prob}
\ee
We can arrange all those $4d^2$ numbers in a column vector, $\bf{P} \in {\cal R}^{4d^2}$, in order to have a geometrical formulation of the problem. For each experimental setting its corresponding joint probability (\ref{prob}) is normalized, hence, $\bf{P}$ must fulfil the four constrains 
\be
  \sum_{k,s=0}^{d-1} P(A_a=k,B_b=s) =1 , \hspace{30pt} a,b=1,2.
\label{normalization}
\ee
Here, we are only interested in correlations that can not be used for instantaneous communication between Alice and Bob. That is, marginal probabilities for one party are independent of the measurement chosen by the other party,
\be
  \sum_{s=0}^{d-1} P(A_a=k,B_1=s) = \sum_{s=0}^{d-1} P(A_a=k,B_2=s) \  .
\label{causality}
\ee
This constraint must hold for each observable, $A_1, A_2, B_1, B_2$, and for each outcome $k=0,\ldots  d-1$, but not all of them are linearly independent. In Lemma 1 (in the appendix) it is proven that the normalization (\ref{normalization}) plus non-signaling (\ref{causality}) conditions form a system of $4d$ linearly independent equations, which will be used later. Then, all possible $\bf{P}$ vectors ---that satisfy these constrains--- belong to an affine space \cite{affine space} of dimension $4d(d-1)$.

In what follows, we characterize all joint probability distributions $P(A_a=k,B_1=s)$ obtainable with local-realistic models. Let $\lambda$ label all possible outcomes of the measurements, that is, the $d^4$ possible values of $(A_1, A_2, B_1, B_2)$. As it is well known, all these models can be written as
\be
  P(A_a=k,B_b=s) = \sum_\lambda\  p_\lambda\, P(A_a=k,\lambda)\, P(B_b=s,\lambda) \hspace{20pt} \mbox{with} \quad p_\lambda \geq 0 \quad \mbox{and} \quad \sum_\lambda p_\lambda = 1 ,
\label{local-realism}
\ee
where $P(A_a=k,\lambda)$ and $P(B_b=s,\lambda)$ take only the values $0,1$. In other words, $P(A_a=k,B_b=s)$ is a convex combination of extreme product probabilities, $P(A_a=k,B_b=s) = \delta_{A_a,k}\ \delta_{B_b,s}$, which when written as column vectors are denoted by ${\bf G}_\lambda$. In our geometric picture,
\be
  {\bf P} = \sum_\lambda\ p_\lambda\, {\bf G}_\lambda\ ,
\ee
is equivalent to say that ${\bf P}$ belongs to the convex hull ($\L$) expanded by the set of vectors ${\bf{ G}_\lambda }$. We say that each one of the $d^4$ vectors ${\bf G}_\lambda$ is a generator of $\L$. Because the number of generators is finite, $\L$ is a convex polytope. It is easy to see that the generators satisfy the normalization (\ref{normalization}) and the non-signaling (\ref{causality}) conditions, and this implies that all the points belonging to the polytope also satisfy them. In Lemma 2 (in the appendix) it is proven that the affine hull \cite{affine hull} expanded by the polytope has dimension $h=4d(d-1)$, just the maximum allowed by these conditions (\ref{normalization}-\ref{causality}) as proved in Lemma 1. It is important to know the dimension of this space, $h$, for what follows.

It is known that every convex polytope is characterized by a unique finite set of inequalities \cite{polytopes}, called \emph{facets}: 
\be
  {\bf P}\in \L \ \Longleftrightarrow \ {\bf X}_j \cdot {\bf P} \leq x_j 
  \hspace{20pt} j=1,2 \ldots 
\label{inequalities}
\ee
Each of these facets fulfil the next two conditions: 
\begin{description}
\item[Condition 1:] Every one of the generators ${\bf G}_\lambda$ must belong either to the half-space ${\bf X}_j \cdot {\bf G}_\lambda < x_j$ or to the hyperplane ${\bf X}_j \cdot {\bf G}_\lambda = x_j$.
\item[Condition 2 (tightness):] Among the generators that belong to the hyperplane ${\bf X}_j \cdot {\bf G}_\lambda = x_j$ there must be $h$ which are affinely independent. Notice that an hyperplane of dimension $h-1$ is completely characterized by $h$ affinely independent points. In our particular case, the null vector does not belong to the polytope, hence the condition of existence of $h$ affinely independent vectors is equivalent to the existence of $h$ linearly independent vectors. 
\end{description}

All these inequalities can be grouped in families of equivalent inequalities. Two inequalities are \emph{equivalent} if we can transform one into the other by composing the following symmetry transformations: 
\begin{description}
\item[Party exchange:] $P(A_a=k,B_b=s) \longrightarrow P(A_b=s,B_a=k)$ or $\langle A_a B_b \rangle \longrightarrow \langle A_b B_a \rangle$.
\item[Observable exchange:] $P(A_a=k,B_b=s) \longrightarrow P(A_{\bar{a}}=k,B_b=s)$ or $\langle A_a B_b \rangle \longrightarrow \langle A_{\bar{a}} B_b \rangle$, where $\bar{1}=2$ and $\bar{2}=1$. 
\item[Relabeling of outcomes:] $P(A_a=k,B_b=s) \longrightarrow P(A_a=\sigma_k,B_b=s)$ or $\langle A_a B_b \rangle \longrightarrow -\langle A_a B_b \rangle$, where $\sigma_k$ is a permutation of the set $\{0,1 \ldots d-1\}$.
\end{description}
Therefore, the obtention of an inequality automatically yields all its family of equivalent ones by transforming it.

Some of these inequalities follow from the fact that the joint probabilities satisfy the normalization and non-signaling conditions, thus quantum joint probabilities never violate them. We call them trivial inequalities. The non-trivial ones are the \emph{tight Bell inequalities} and could be violated by quantum correlations. 

Non-tight Bell inequalities are the ones that fulfil Condition 1 but not Condition 2. They are worse detectors of non-local-realistic correlations because they could be modified in order to detect the same points as before plus additional ones. In the next section we show that the CGLMP-inequality is tight.

\section{The CGLMP-inequality} 

In this section we will show that the CGLMP-inequality satisfy condition 1 and condition 2 stated in the previous section. If we use the symbol $\doteq$ to denote equality modulus $d$, the CGLMP-inequality for any value of $d$ can be written as
\bea
  I_d({\bf P}) = \sum_{k=0}^{\lfloor d/2-1 \rfloor} \left( 1 - \frac{2k}{d-1} \right)   
\bigg[ P(A_1 - B_1 \doteq k) - P(A_1 - B_1 \doteq -k-1) + P(A_1 - B_2 \doteq -k)  - P(A_1 - B_2 \doteq k+1) \cr + P(A_2 - B_1 \doteq -k-1) - P(A_2 - B_1 \doteq k) + P(A_2 - B_2 \doteq k) - P(A_2 - B_2 \doteq -k-1)  \bigg] \leq 2 \ , 
\label{inequality}\eea 
where $\lfloor x \rfloor$ denotes the largest integer less or equal than $x$. Notice that for $d=2$ (\ref{inequality}) is equivalent to the famous CHSH-inequality \cite{CHSH}. The main result of this paper is that, {\em for all $d$, the CGLMP-inequality is tight.} In order to prove it, we need to show that both condition 1 and 2 are satisfied by (\ref{inequality}). In \cite{Gisin} the authors showed that condition 1 is satisfied by the inequality. In what follows, we will reproduce their proof for the sake of completeness.

\bigskip

\noindent {\bf Proof of Condition 1:} Let us start by defining the four variables
\be
r= A_1 - B_1 +\dot{d}\ ,\hspace{20pt}  s= -A_1 + B_2+\dot{d}\ ,\hspace{20pt} t=-A_2 + B_1 -1+\dot{d}\ ,\hspace{20pt} u= A_2 - B_2+\dot{d}\ , 
\label{rstu}\ee
were $\dot{d}$ denotes a multiple of $d$ (possibly different for each variable) that has to be added in order to have that 
\be
  - \left\lfloor \frac{d}{2} \right\rfloor \leq r,s,t,u \leq \left\lfloor \frac{d-1}{2} \right\rfloor 
\label{restriction}\ee
is satisfied. Recalling the definition of $\doteq$ and $\dot{d}$, it is easy to see that this new variables fulfil the constrain
\be
  r+s+t+u \doteq -1 \ .
\label{-1}\ee 
Using these variables we have defined, the inequality (\ref{inequality}) can be written as
\bea
  I_d({\bf P}) = \sum_{k=0}^{\lfloor d/2-1 \rfloor} \left( 1 - \frac{2k}{d-1} \right)   
\bigg[ &P(r=k) - P(r=-k-1) + &P(s=-k)  - P(s=k+1) \cr + &P(t=-k-1) - P(t=k) + &P(u=k) - P(u=-k-1)  \bigg] \leq 2 \ . 
\label{Gropius} \eea
Where, because we have defined $r,s,t,u$ such that (\ref{restriction}) is satisfied, the $\doteq$ symbols in (\ref{inequality}) can be substituted by simple equalities. After some algebra, the value of $I_d$ when applied to the generators, ${\bf G}_\lambda$, yields 
\be
  I_d({\bf G}_\lambda) = f(r) + f(s) + f(t) + f(u) \hspace{40pt} 
  f(x)= \left\{ \matrix{
		  -\frac{2x}{d-1}+1  &,   &x\geq 0 \cr 
		  -\frac{2x}{d-1}-\frac{d+1}{d-1} &,  &x<0
                } \right. \hspace{20pt} ,
\label{I_d}\ee
where the values of $r,s,t,u$ are the ones corresponding to each $\lambda$. Now, to prove Condition 1, we look at the four cases that appear when considering the different sings the variables $r,s,t,u$ can have: 
\begin{description}
\item[Case 1.] $r,s,t,u$ are all positive. Then, (\ref{restriction}) and (\ref{-1}) imply that $r+s+t+1=d-1$, which when inserted into (\ref{I_d}) gives $I_d=2$.
\item[Case 2.] Three of the variables $r,s,t,u$ are positive and one is strictly negative. Then, (\ref{restriction}) and (\ref{-1}) imply that either $r+s+t+u=d-1$ which gives us $I_d=-2/(d-1)$, or, $r+s+t+u=-1$ which give us $I_d=2$.
\item[Case 3.] Two of the variables $r,s,t,u$ are positive and the other two are strictly negative. Then, (\ref{restriction}) and (\ref{-1}) imply that $r+s+t+u=-1$, giving $I_d=-2/(d-1)$. 
\item[Case 4.] One of the variables $r,s,t,u$ is positive and the other three are strictly negative. Then, (\ref{restriction}) and (\ref{-1}) imply that either $r+s+t+u=-1$ or $r+s+t+u=-d-1$ happens. Then, the two possible values for (\ref{I_d}) are either $I_d=-2(d+1)/(d-1)$ or  $I_d=-2/(d-1)$.
\item[Case 5.] $r,s,t,u$ are all strictly negative. Then, (\ref{restriction}) and (\ref{-1}) imply $r+s+t+u=-d-1$ and then $I_d=-2(d-1)/(d-1)$.
\end{description}
We have seen that for all the generators, ${\bf G}_\lambda$, the value $I_d({\bf G}_\lambda)$ is equal or less than $2$, which completes the proof of condition 1.  

Until now, it has been proven that the CGLMP-inequality is a Bell inequality, that is, any set of joint probabilities that violates it is not inside the polytope. In what follows, we will see that this inequality (\ref{inequality}) is a facet of the polytope, or in other words, it is as tight as it can be.

\bigskip

\noindent {\bf Poof of Condition 2 (tightness):} As we have seen in the proof of Condition 1, the generators contained in the hyperplane ---that is, those that satisfy $I_d(G_\lambda)=2$--- are (i) the ones having the four variables $r,s,t,u$ positive, and (ii) the ones having three variables positive and one strictly negative that satisfy $r+s+t+u=-1$. Now, we will demonstrate that among these generators there are $h$ which are linearly independent. First notice that all the generators can be written as
\be
{\bf G}_{(A_1,A_2,B_1,B_2)} = \ket{A_1,B_1} \oplus \ket{A_1,B_2} \oplus \ket{A_2,B_1} \oplus \ket{A_2,B_2} \ , 
\label{ket}\ee
where $\ket{A,B}$ stands for $\ket{A \bmod d}\otimes\ket{B \bmod d}$, and $\ket{n}$ is a $d$-dimensional vector with a $1$ in the $n^{\text{th}}$ component and $0$s in the rest\footnote{To avoid confusion we remark that the order of the components in (\ref{ket}) is not the same than in (\ref{generators}), in Lemma 2.}. Of course, every one of the four spaces in the direct sum has dimension $d^2$. Using the variables defined in (\ref{rstu}) we can also write (\ref{ket}) as 
\be
\ket{A_1,A_1-r} \oplus \ket{A_1,A_1+s} \oplus \ket{A_1-r-t-1,A_1-r} \oplus \ket{A_1+s+u,A_1+s} \ .  
\label{Proust}\ee
Because we are only interested in the linear dependence properties of these vectors, we can apply to them a linear transformation that preserves orthogonality. Thus, (\ref{Proust}) can be transformed into 
\be
\ket{A_1,r} \oplus \ket{A_1,s} \oplus \ket{A_1-r,t} \oplus \ket{A_1+s,u} \ ,
\label{Wittgenstein}\ee
by applying the permutation $\sum_{A,k=0}^{d-1} \left( \ket{A,k}\!\bra{A,A-k} \otimes  \ket{A,k}\!\bra{A,A+k} \otimes \ket{A,k}\!\bra{A-k-1,A} \otimes  \ket{A,k}\!\bra{A+k,A} \right)$. which is an orthogonal matrix. The construction of the set of $4d(d-1)$ linearly independent vectors is done step by step. In each step, $4 d$ vectors which are linearly independent among them and among the previously introduced vectors, are added. The following two examples show how to introduce a set of $4 d$ vectors, for two particular cases, that will be used later for the general case.

\begin{description}
\item[Example 1:] Imagine that only vectors of the form (\ref{Wittgenstein}) having the values of $r,s,t,u$ different from $a$ have been introduced. Then, the following $4d$ vectors
\bea \matrix{
  \ket{A,a} \oplus \ket{A,b_1} \oplus \ket{A-a,b_2} \oplus \ket{A+b_1,b_3} \cr 
  \ket{A,b_3} \oplus \ket{A,a} \oplus \ket{A-b_3,b_1} \oplus \ket{A+a,b_2} \cr 
  \ket{A,b_2} \oplus \ket{A,b_3} \oplus \ket{A-b_2,a} \oplus \ket{A+b_3,b_1} \cr 
  \ket{A,b_1} \oplus \ket{A,b_2} \oplus \ket{A-b_1,b_3} \oplus \ket{A+b_2,a} }
  \quad \mbox{for}\quad  A=0,1\ldots d-1,
\label{i1}\eea 
where $b_1,b_2,b_3 \neq a$, can be added to the set, with the certainty that they are linearly independent form the previously introduced ones ---since they did not have components like $\ket{A,a}$ in any of the four orthogonal subspaces---, and that they are also linearly independent among them, because $\ket{A,a}$ is, for each vector, in a different orthogonal subspace. Notice that argument is valid for all the values of $b_1,b_2,b_3$ different from $a$, independently from the fact that they have appeared in a previously introduced vector, or not.

\item[Example 2:] Imagine that only vectors of the form (\ref{Wittgenstein}) having the values of $r,s,t,u$ different from $a$ have been introduced. Then, the following $4d$ vectors
\bea \matrix{
  \ket{A,a} \oplus \ket{A,a} \oplus \ket{A-a,b_1} \oplus \ket{A+a,b_2} \cr 
  \ket{A,a} \oplus \ket{A,b_1} \oplus \ket{A-a,a} \oplus \ket{A+b_1,b_2} \cr 
  \ket{A,a} \oplus \ket{A,b_1} \oplus \ket{A-a,b_2} \oplus \ket{A+b_1,a} \cr 
  \ket{A,b_1} \oplus \ket{A,a} \oplus \ket{A-b_1,a} \oplus \ket{A+a,b_2} }
  \quad \mbox{for}\quad  A=0,1\ldots d-1
\label{i2}\eea 
can be added to the set, with the certainty that they are linearly independent form the previously introduced ones, for the same reason than in example 1. To see that they are also linearly independent among them, we can observe that for a fixed value of $A$, the projection of these four vectors onto the subspace $\proj{A,a} \oplus \proj{A,a} \oplus \proj{A-a,a} \oplus \proj{A+b_1,a}$ contains the minor 
\be \left[\matrix{ 
1 & 1 & 0 & 0 \cr
1 & 0 & 1 & 0 \cr
1 & 0 & 0 & 1 \cr
0 & 1 & 0 & 0 } \right]\ ,
\ee
which has a non-zero determinant. The argument is also valid for all the values of $b_1,b_2$ different from $a$, independently from the fact that they have appeared in a previously introduced vector.
\end{description}

Now, we proceed to construct the set of $4d(d-1)$ linearly independent vectors. As we said, we will do this operation in $d-1$ steps. In each of these steps, $4d$ vectors are introduced in the same way than in examples 1 and 2, but, with different values for $a,b_1,b_2,b_3$ in each step. To simplify this operation, we study separately four cases, namely: $d = 4e$, $4e+1$, $4e+2$ and $4e+3$, where $e$ is an arbitrary integer.
\begin{description}

\item[Case $d = 4 e$.] In this case the constrain (\ref{restriction}) is $-2e\leq r,s,t,u \leq 2e-1$. For the first $2e-1$ steps, we introduce vectors of the form (\ref{Wittgenstein}) with $r,s,t,u \geq 0$, for which (\ref{-1}) is $r+s+t+u=4e-1$. In each step, the scheme of example 1 (\ref{i1}) is used with the following values for $a,b_1,b_2,b_3$ of:
\be
\matrix{a\cr b_1\cr b_2\cr b_3}\ = \quad \matrix{e-1\cr e\cr e\cr e}\ , \quad \matrix{e+1\cr e-1\cr e-1\cr e}\ , \quad \matrix{e-2\cr e+1\cr e\cr e}\ , \quad \matrix{e+2\cr e-2\cr e-1\cr e}\ , \quad \matrix{e-3\cr e+2\cr e\cr e}\ , \quad \matrix{e+3\cr e-3\cr e-1\cr e}\ ,  \ldots \quad \matrix{e-k\cr e+k-1\cr e\cr e}\ , \quad \matrix{e+k\cr e-k\cr e-1\cr e}\ \ldots \quad \matrix{0\cr 2e-1\cr e\cr e} 
\ee
Notice that in each step ---reading from left to right--- the top value of each column appears for the first time, hence the scheme of example 1 can be applied in every step. After this operation we have a set of $4d(2e-1)$ vectors. For the next $2e$ steps, we introduce vectors of the form (\ref{Wittgenstein}) having three of the variables $r,s,t,u$ positive and one strictly negative that fulfil $r+s+t+u=-1$. We also use the scheme of the first example but with the following values for $a,b_1,b_2,b_3$:
\be
\matrix{a\cr b_1\cr b_2\cr b_3}\ = \quad \matrix{-1\cr 0\cr 0\cr 0}\ , \quad \matrix{-2\cr 1\cr 0\cr 0}\ , \quad \matrix{-3\cr 2\cr 0\cr 0}\ \ldots \quad \matrix{-k\cr k-1\cr 0\cr 0}\ \ldots \quad \matrix{-2e\cr 2e-1\cr 0\cr 0}
\ee
Notice that as before, the top value of every column appears also for the first time. After this, we have constructed a set of $4d(d-1)$ linearly independent vectors. This finishes the proof for the case $d=4e$.

\item[Case $d = 4 e +1$.] In this case (\ref{restriction}) is  $-2e\leq r,s,t,u \leq 2e$. For the first $2e$ steps, we introduce vectors with $r,s,t,u \geq 0$ and $r+s+t+u=4e$ in the following order:
\bea
&\matrix{b_1\cr b_2\cr a\cr a}\ = \quad \matrix{e+1\cr e+1\cr e-1\cr e-1}\ , \quad \matrix{e-1\cr e+1\cr e\cr e}\ \cr
&\matrix{a\cr b_1\cr b_2\cr b_3}\ = \quad \matrix{e+2\cr e-1\cr e-1\cr e}\ , \quad \matrix{e-2\cr e+2\cr e\cr e}\ \ldots \quad \matrix{e+k\cr e-k+1\cr e-1\cr e}\ , \quad \matrix{e-k\cr e+k\cr e\cr e}\ \ldots \quad \matrix{2e\cr 1\cr e-1\cr e}\ , \quad \matrix{0\cr 2e\cr e\cr e}
\eea
The first two steps have to be done following the second example, because each one has two times a value that appears for the first time. For all the rest, the top value appears for the first time, thus the first example can be applied as model. After this operation we have introduced $4d\,2e$ vectors in our set. For the next $2e$ steps, we introduce vectors having three of the variables $r,s,t,u$ positive and one strictly negative and $r+s+t+u=-1$. We also use the scheme of example 1 in the next steps:
\be
\matrix{a\cr b_1\cr b_2\cr b_3}\ = \quad \matrix{-1\cr 0\cr 0\cr 0}\ , \quad \matrix{-2\cr 1\cr 0\cr 0}\ \ldots \quad \matrix{-k\cr k-1\cr 0\cr 0}\ \ldots \quad \matrix{-2e\cr 2e-1\cr 0\cr 0}
\ee
We have constructed a set of $4d(d-1)$ linearly independent vectors, which finishes the proof for the case $d=4e+1$.
 
\item[Case $d = 4 e +2$.] In this case $-(2e+1)\leq r,s,t,u \leq 2e$. For the first $2e$ steps, we introduce vectors which have $r,s,t,u \geq 0$ and $r+s+t+u=4e+1$, using the scheme of example 1, in the following order:
\be
\matrix{a\cr b_1\cr b_2\cr b_3}\ = \quad \matrix{e+1\cr e\cr e\cr e}\ , \quad \matrix{e-1\cr e+1\cr e+1\cr e}\ , \quad \matrix{e+2\cr e-1\cr e\cr e}\ , \quad \matrix{e-2\cr e+2\cr e+1\cr e}\ \ldots \quad \matrix{e+k\cr e-k+1\cr e\cr e}\ , \quad \matrix{e-k\cr e+k\cr e+1\cr e}\ \ldots \quad \matrix{2e\cr 1\cr e\cr e}\ , \quad \matrix{0\cr 2e\cr e+1\cr e} 
\ee
For the next $2e+1$ steps, we introduce vectors having three of the variables  $r,s,t,u$ positive and one strictly negative and $r+s+t+u=-1$, using the scheme of example 1, in the following order:
\be
\matrix{a\cr b_1\cr b_2\cr b_3}\ = \quad \matrix{-1\cr 0\cr 0\cr 0}\ , \quad \matrix{-2\cr 1\cr 0\cr 0}\ \ldots \quad \matrix{-k\cr k-1\cr 0\cr 0}\ \ldots \quad \matrix{-(2e+1)\cr 2e\cr 0\cr 0}
\ee
We have constructed a set of $4d(d-1)$ linearly independent vectors, which finishes the proof for the case $d=4e+2$.

\item[Case $d = 4 e +3$.] In this case $-(2e+1)\leq r,s,t,u \leq 2e+1$. For the first $2e+1$ steps, we introduce vectors with $r,s,t,u \geq 0$ and $r+s+t+u=4e+2$ in the following order:
\bea
&\matrix{a\cr b_1\cr a\cr b_2}\ = \quad \matrix{e+1\cr e\cr e+1\cr e}\ \cr
&\matrix{a\cr b_1\cr b_2\cr b_3}\ = \quad\matrix{e-1\cr e+1\cr e+1\cr e+1}\ , \quad \matrix{e+2\cr e-1\cr e+1\cr e}\ , \quad \matrix{e-2\cr e+2\cr e+1\cr e+1}\ \ldots \quad \matrix{e+k\cr e-k+1\cr e+1\cr e}\ , \quad \matrix{e-k\cr e+k\cr e+1\cr e+1}\ \ldots \quad \matrix{2e+1\cr 0\cr e+1\cr e} 
\eea
The first step has to be done following example 2. For all the rest the scheme of the first example can be applied. For the next $2e+1$ steps, we introduce vectors having three of the variables $r,s,t,u$ positive and one strictly negative and $r+s+t+u=-1$, using the scheme of the first example:
\be
\matrix{a\cr b_1\cr b_2\cr b_3}\ = \quad \matrix{-1\cr 0\cr 0\cr 0}\ , \quad \matrix{-2\cr 1\cr 0\cr 0}\ \ldots \quad \matrix{-k\cr k-1\cr 0\cr 0}\ \ldots \quad \matrix{-(2e+1)\cr 2e\cr 0\cr 0}
\ee
We have constructed a set of $4d(d-1)$ linearly independent vectors, which finishes the proof for the last remaining case. 
\end{description}

Thus, we have shown that the CGLMP-inequality satisfies not only condition 1, but also condition 2, and therefore it is a tight Bell inequality for all values of $d$.

\section{Complete characterisation for generalized 3-outcome correlation functions}

In this section we give a way to simplify the data extracted from bipartite measurements, that is a generalization of the two-outcome correlation functions. With this simplification numerical work becomes possible and some results are obtained.

Let us concentrate on the case $d=2$, and let us label the two outcomes of all the observables by the numbers $1$ and $-1$, that is $A_a,B_b=1,-1$. The correlation function of the variables $A_a$ and $B_b$ is defined as the average of their product, $\langle A_a B_b \rangle$, for different realizations of the experiment. Now, instead of dealing with the 16-dimensional vectors with components
\be
  P(A_a=k,B_b=s) \hspace{30pt} k,s=1,-1 \hspace{30pt} a,b=1,2  
\label{16-vector} \ee 
(like in section II and III), we could deal with the 4-dimensional vectors that have the correlation functions 
\be 
  \langle A_1 B_1 \rangle,\ \langle A_1 B_2 \rangle,\ \langle A_2 B_1 \rangle,\ \langle A_2 B_2 \rangle  
\label{4-vector} \ee
as components. To transform a vector from the 16-dimensional space to its corresponding vector in the 4-dimensional space (\ref{4-vector}), we have to perform the projection
\be
  \langle A_a B_b \rangle = P(A_a=1,B_b=1) + P(A_a=-1,B_b=-1) - P(A_a=1,B_b=-1) - P(A_a=-1,B_b=1) \ ,
\label{projection} \ee
in each of the four subspaces $a,b=1,2$. It is known that the projection of a convex polytope is always a convex polytope \cite{polytopes}. Therefore, the set of vectors of correlation functions (\ref{4-vector}) achievable with local-realistic models is also a polytope, characterized by a new set of Bell inequalities. The CHSH-inequality \cite{CHSH} 
\be
  \langle A_1 B_1 \rangle + \langle A_1 B_2 \rangle + \langle A_2 B_1 \rangle - \langle A_2 B_2 \rangle \leq 2 \ , 
\label{CHSH} \ee
is an instance. Even more, it was proven in \cite{Fine}, that all the non-trivial facets of this polytope are equivalent to the CHSH-inequality. Hence, the satisfiability of this inequality ---and its equivalent ones--- by a set of correlation functions is a necessary and sufficient condition for the existence of a local-realistic model giving these correlation functions. For the setting consisting of an arbitrary number of parties each possessing two dichotomic observables, all the Bell inequalities for correlation functions have also been obtained \cite{Werner}. 

Dealing with correlation functions is much more simple than with joint probabilities, although in general, after doing the projection (\ref{projection}), some non-local-realistic joint probabilities can be projected into local-realistic correlation functions. It was proven in \cite{Fine} that in the setting of the CHSH-inequality ($d=2$) this do not happen, or in other words, correlation functions contain the same information that joint probabilities concerning the existence of a local-realistic model accounting for them. In what follows we show a generalization of the correlation functions that can be used with observables of more than two outcomes: Instead of using the $4d^2$-dimensional vectors of joint probabilities (\ref{prob}), let us deal with the $4d$-dimensional vectors with components:
\be
  P(A_a-B_b \doteq n) = \sum_{j=0}^{d-1} P(A_a = n+j\bmod d, B_b=j) \ .
\label{nomad} \ee
We say that the $P(A_a-B_b \doteq n)$'s generalize the idea of correlation functions because they are also a projection of the joint probabilities and, for $d=2$ (\ref{nomad}) contains the same information than (\ref{projection}). Notice that this description is equivalent to say that only the information of the variables $r,s,t,u$ (section III) is used. Because the CGLMP-inequality (\ref{Gropius}) can be written in terms of only $r,s,t,u$, the projection (\ref{nomad}) leaves it unchanged, and this implies that it is also a facet of the new polytope that yields the projection (\ref{nomad}). This polytope is the convex hull of the $d^3$ generators
\be
\ket{A_1-B_1} \oplus \ket{A_1-B_2} \oplus \ket{A_2-B_1} \oplus \ket{A_2-B_2} 
\hspace{40pt} A_1,A_2,B_1,B_2 = 0,1 \ldots d-1.
\ee
We have done an algorithm that finds all the facets of this polytope for any $d$. Contrary to what happens with the polytope of general joint probabilities, in this case the program runs quite fast for $d=3$, and it turns out that, all non-trivial inequalities are equivalent to the CGLMP-inequality:

{\bf Result:} \emph{The satisfyability of the CGLMP-inequality for $d=3$ ---and its equivalent ones--- by a set of generalized correlation functions (\ref{nomad}) is a necessary and sufficient condition for the existence of a local-realistic model accounting for them.}

Hence, we have a simple complete characterization of all local-realistic generalized correlation functions of three outcomes. For $d=4$ the program takes too much time for an exhaustive search, but we have found that, there are at least three non-trivial inequalities which are not equivalent. This makes natural the assumption that for any $d$ larger than three there is not a unique class of inequalities.

\section{Conclusions}

In this paper we have proven that the CGLMP-inequality is tight, by taking a geometric point of view. This is the first time that the tightness of a Bell inequality for an arbitrary number of outcomes is proven. There are not much criteria for evaluating the power of a Bell inequality, and tightness is a very objective one. 

We give a possible generalization of correlation functions that can deal with more than two-outcome experiments. With this definition a complete characterization of the polytope for $d=3$ is numerically possible, and we have found that, 
all non-trivial facets are equivalent to the CGLMP-inequality. That is, for $d=3$, the satisfyability of this inequality ---and its equivalent ones--- by a set of generalized correlation functions is a necessary and sufficient condition for the existence of a local-realistic model accounting for them.

The fact that this inequality is not maximally violated by the maximally entangled state, for $d\geq 3$, could be due to the usage of only two Von Newman observables per party. Perhaps, increasing the number of observables per party or allowing for POVMs could remove this bizarre situation, although this is an open problem.

\bigskip

The author is most grateful to J. I. Latorre for its help in the design of the algorithm and many interesting discussions. The author also thanks A. Ac\'{\i}n, E. Jan\'e and E. Rico for their useful comments and suggestions on the final manuscipt. This work is financially supported by the following projects: AEN99-0766, 1999SGR-00097, IST-1999-11053. G.V. and the grant 2002FI-00373 UB.

\section{Appendix}

Here there are the two lemmas that have not been written in the main body of the paper:

\begin{description} 

\item[Lemma 1:] Here, we prove that the non-signaling (\ref{normalization}) and normalization (\ref{causality}) conditions give us a set of $4d$ linearly independent equations. 

To achieve this, we construct a set of $4d$ equations by adding one at every step which contains a coordinate that the previously introduced equations did not contain. First, we introduce the following equations:
\be
  \sum_{s=0}^{d-1} P(A_1=k,B_1=s) = \sum_{s=0}^{d-1} P(A_1=k,B_2=s) \hspace{40pt} k=0,1,\ldots d-2 ,
\label{eqs} \ee
taken from the set (\ref{causality}) ---notice that the equation corresponding to $k=d-1$ is not taken---. It is easy to see that for a particualr value  $k=k_0$ the coordinate $P(A_1=k_0,B_1=d-1)$ appears only in one of the equations ---the one concerning $A_1=k_0$--- and therefore (\ref{eqs}) is a set of $d-1$ linearly independent equations. Second, we introduce the set of equations obtained by substituting $A_1$ by $A_2$ in (\ref{eqs}). As before, for every particular value $k=k_0$ the coordinate $P(A_2=k_0,B_1=d-1)$ appears only in one of the equations introduced until now. Next, we do the same but with $B_1$ and $B_2$. After these operations, we have constructed a set of $4(d-1)$ linearly independent equations. Finally, we add the four normalization constrains (\ref{normalization}). Notice that in each of these four equation a coordinate like $P(A_a=d-1,B_b=d-1)$ appears for the first time. This happens because we have excluded the four equations corresponding to the value $k=d-1$ from the non-signaling condition. It can be seen that these four equations are linear combinations of the ones in the set we have constructed. Hence we have proved that among the equations (\ref{normalization}-\ref{causality}) there are only $4 d$ which are linearly independent.

\item[Lemma 2:]
Here, we prove that the affine hull of the polytope has dimension $4d(d-1)$. 

By construction all the generators satisfy the normalization (\ref{normalization}) and the non-signaling (\ref{causality}) conditions, hence the whole polytope belong to the affine space found in Lemma 1. A generic generator with components $\delta_{A_a,k}\, \delta_{B_b,s}$ can be written in this way: 
\be
  {\bf G}_{(A_1, A_2, B_1, B_2)} = \left( \ket{A_1} \oplus \ket{A_2} \right) \otimes \left( \ket{B_1} \oplus \ket{B_2} \right) .
\label{generators}
\ee
Where the symbol $\ket{A}$ represents a $d$-dimensional column vector with a $1$ in the $A^{\text th}$ component and a $0$ in the rest. Let us concentrate for a while on the set of vectors 
\be
  \left( \ket{A_1} \oplus \ket{A_2} \right) \in {\cal R}^{2d} \hspace{30pt} A_1, A_2 = 0,1 \ldots d-1 . 
\label{petits} \ee 
It is easy to see that the following $2 d-1$ vectors are linearly independent: 
\be
  \left( \ket{0} \oplus \ket{0} \right),\  \left( \ket{0} \oplus \ket{1} \right),\  \left( \ket{0} \oplus \ket{2} \right)\ \ldots\ \left( \ket{0} \oplus \ket{d-1} \right),\ \left( \ket{1} \oplus \ket{d-1} \right),\ \left( \ket{2} \oplus \ket{d-1} \right)\ \ldots\ \left( \ket{d-1} \oplus \ket{d-1} \right) .
\label{A vectors}
\ee
Thus, we can say that in the set (\ref{petits}) there are at least $2d-1$ linearly independent vectors, or in other words, the linear span of this set has dimension larger or equal than $2d-1$. The generators (\ref{generators}) are tensor products of vectors of the form (\ref{petits}). This implies that the linear span of the generators has dimension larger or equal than $(2d-1)^2$. Because the null vector cannot be obtained by any affine combination \cite{affine hull} of the generators, their affine hull has one dimension less that their linear span. This means that the dimension, $h$, of this affine space is larger or equal than $4d(d-1)$. But we know from Lemma 1 that the polytope belongs to an affine space of dimension $4d(d-1)$, which implies that the dimension of its affine hull, $h$, cannot be larger than $4d(d-1)$. Then, the only possibility is that the dimension of the affine space generated by the polytope is exactly $h=4d(d-1)$, which concludes the proof.

\end{description}

\end{document}